\documentclass[11pt,twoside]{article}
\usepackage{asp2006}
\usepackage{psfig}
\usepackage{epsf}
\usepackage{graphics}
\usepackage{lscape}
\def\edcomment#1{\iffalse\marginpar{\raggedright\sl#1\/}\else\relax\fi}
\reversemarginpar

\begin{document}
\title{Minima of Solar Cycles 22/23 and 23/24 as Seen in \\
SOHO/CELIAS/SEM Absolute Solar EUV Flux}
\author{Leonid V. Didkovsky, Darrell L. Judge, Seth R. Wieman}
\affil{University of Southern California, Space Sciences Center, Los Angeles, CA 90089-1341,
USA}
\author{Don McMullin}
\affil{The Space Systems Research Corp., Alexandria, VA 22314, USA}

\begin{abstract}
Verified and updated calibrated absolute solar flux in the He~II 30.4~nm spectral band-pass as measured by the Solar EUV Monitor (SEM) allows us to study variations of the solar EUV irradiance near the minima of Solar Cycles 22/23 and 23/24. Based on eight (1996 to 2007) NASA sounding rocket flights, a comparison of SEM data with the measurements from three independent EUV instruments was performed to verify and confirm the accuracy of the published SEM data. SEM calibrated data were analyzed to determine and compare minima for solar cycles 22/23 and 23/24. The minima points were calculated using SEM first order daily averaged flux smoothed by a running mean (RM) filter with the window of averaging equal to 365 days. These minima occurred on June 2, 1996 (22/23) and November 28, 2008 (23/24). The 23/24 minimum showed about 15\% lower EUV flux in the 30.4 nm band-pass than the 22/23 minimum. The 365-day RM curve around the 23/24 minimum has significant asymmetry (fast decrease of the EUV flux to the minimum and a long, near-horizontal profile after the minimum). This profile is quite different from the much faster and symmetrical change of the flux around the 22/23 minimum. SEM flux was compared with both high spectral resolution (0.1~nm) Mg~II index calculated from the Solar Radiation and Climate Experiment (SORCE) using the Solar Stellar Irradiance Comparison Experiment (SOLSTICE) data and with the NOAA composite Mg~II index spectrum.
\end{abstract}

\vspace{-0.5cm}
\section{Introduction}

SOHO/CELIAS SEM \citep{Hovestadt95} is a highly stable and accurate solar EUV spectrometer. The stability of the spectrometer is provided by a transmission diffraction grating based on thin gold bars \citep{Schattenburg90}. The two (plus and minus) first order bands are centered about the He II (30.4~nm) EUV spectral line. The high accuracy of SEM EUV measurements is achieved by correcting for changes of the flight SEM sensitivity to the solar EUV irradiance based on comparisons of the flight data with the data from a number of NASA sounding rocket calibration flights. The sounding rocket flights have provided solar measurements to correct for these changes in sensitivity, which are due to minor degradation of SOHO SEM's thin film Al filters \citep{Judge98}. A model of such time dependent degradation was created based on the data from the three earliest SOHO sounding rocket under-flights (1996 to 2000). The model was confirmed by the following four under-flights (2001 to 2006). This comparison of the SOHO SEM solar flux measured during the sounding rocket under-flights showed that the absolute solar flux is accurate \citep{Judge08} to within $\pm 5\%$ of the measured flux.

The long, (more than 13.8 years) and practically uninterrupted absolute EUV flux in the mean (plus and minus) first order band around the He II spectral line is ideally suited for an analysis of the 22/23 and 23/24 solar minima.

\section{How the EUV Flux is Calculated}
SEM absolute EUV flux $F$ is calculated from the effective counts (${DN - background}$) measured by each channel's electrometer.
\begin{equation}
F = {k(\lambda,degrad)} \frac {DN-background}{A\frac{\sum{_{\lambda1}^{\lambda2}\epsilon (\lambda)\Phi_{S22}(\lambda) \Delta\lambda}}{\sum{_{\lambda1}^{\lambda2} \Phi_{S22}(\lambda) \Delta\lambda}}f_{1AU} f_{degrad}(\lambda) f_{atm}(\lambda)}
\end{equation}
The spectral distribution of the solar irradiance ${\Phi_{S22}(\lambda)}$ is given by the SOLERS-22 model, a composite spectrum \citep{Woods98}. Efficiency ${\epsilon(\lambda)}$ is the channel's responsivity profile determined during the NIST calibration. The entrance slit aperture is ${A}$. The coefficient ${k(\lambda,degrad)}$ corrects for the higher order contribution, e.g., for relatively strong spectral lines in the 17~nm band. The correction for the variable distance from the Sun is provided by the ${f_{1AU}}$ coefficient. Degradation of the Al filters is given by ${f_{degrad}(\lambda)}$ for the SEM on SOHO. Correction for the transmission of the Earth's atmosphere for the sounding rocket flights is provided by the ${f_{atm}(\lambda)}$.

\section{Verification of the SEM EUV Absolute Flux}
EUV absolute flux from SOHO/CELIAS SEM measurements were verified by comparing them with the EUV measurements provided by our NASA sounding rocket flights. Table 1 summarizes the results of this comparison.
\begin{table}[!ht]
\label{Verification}
\caption{A comparison of SOHO/SEM and sounding rocket EUV flux measurements. The flux units are $1.0\times 10^{10}{ph/cm^{2}/s}$.}
\smallskip
\begin{center}
{\small
\begin{tabular}{lllllll}
\tableline
\noalign{\smallskip}
Date&NASA&SOHO/&SEM&Ne RGIC& EVE/& Ratio\\
&rocket&SEM flux&clone &flux & ESP&to\\
&flight & &flux&&flux &SOHO/\\
& & & & & & SEM\\
\noalign{\smallskip}
\tableline
\noalign{\smallskip}
06/26/1996&36.147&1.21&&1.15&&\textbf{0.95}\\
08/11/1997&36.164&1.42&1.28&1.36& &\textbf{0.96}\\
08/18/1999&36.181&2.22&2.09&2.24& &\textbf{1.01}\\
08/06/2002&36.202&2.28&2.29&2.43& &\textbf{1.06}\\
12/05/2003&36.211&1.78&1.75&1.67& &\textbf{0.94}\\
08/03/2005&36.227&1.57&1.53&1.52& &\textbf{0.96}\\
11/07/2006&36.236&1.26&1.20&1.22& &\textbf{0.97}\\
04/14/2008&36.240&0.953 & & &0.859&\textbf{0.90}\\
\noalign{\smallskip}
\tableline\
\end{tabular}
}
\end{center}
\end{table}
The daily averaged EUV flux from SOHO/CELIAS SEM (third column) was compared with the flux from the SEM clone, a prototype of the flight SEM instrument (fourth column), and with the Ne Rare Gas Ionization Cell (RGIC) which is sensitive to the solar EUV irradiance in the wavelengths from 5 to 57.5~nm. This extended RGIC bandpass was transferred to the SEM bandpass of 26 to 34~nm (fifth column) using the solar model SOLERS-22 \citep{Woods98}. The sixth column shows a flux data point from the Extreme ultraviolet SpectroPhotometer (ESP) \citep{D09} which is an advanced version of SEM and part of the Solar Dynamics Observatory (SDO) EUV Variability Experiment (EVE) \citep{Woods09a}. The last column shows the ratio between the RGIC, ESP  and SOHO SEM. The mean ratio for all compared flights is in the range of $\pm 5\%$. The daily averaged EUV flux from SOHO/SEM together with the flux points from the SEM clone, RGIC, and ESP on the sounding rockets is shown in Figure 1.
\setcounter{figure}{0}
 \begin{figure}[!ht]
 \plotone {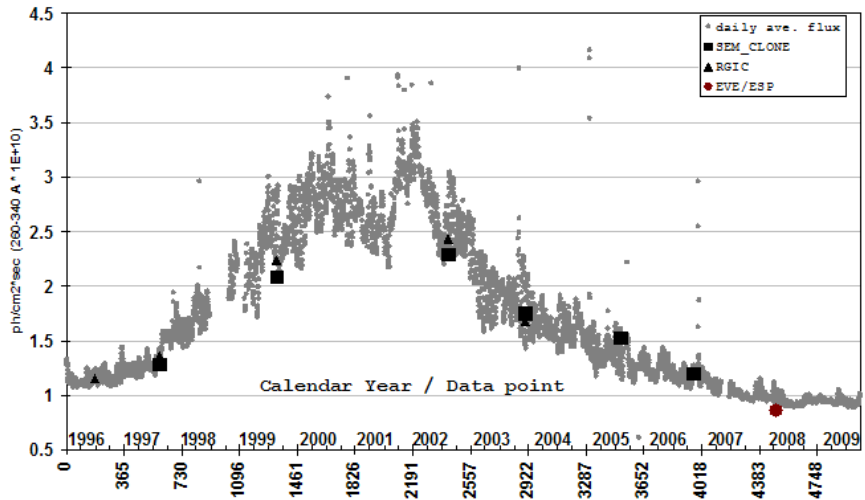}
 \caption{Daily averaged SOHO/SEM flux points (grey) from 1 January 1996 through 28 September 2009 with over-plotted sounding rocket measurements from SEM clone (squares), RGIC (triangles), and SDO/EVE/ESP (circle). The mean scatter of the sounding rocket points around the SOHO/SEM points is $\pm 5\%$.}
 \end{figure}
The sounding rocket points (black squares and triangles are for the SEM clone and RGIC, respectively, and the circle is for the rocket ESP) match the SEM daily averaged points (grey) within $\pm 5\%$. A portion of this range is related to a shorter integration time (a couple of minutes) for the rocket's near-apogee data points compared to the SEM daily averaged points.

\section{A Comparison of SEM Flux with Mg II Solar Index Near 22/23 and 23/24 Solar Minima }

To compare SEM EUV flux centered around the He II (30.4~nm) spectral line with Mg II index, SEM flux was scaled as $SEM/1.1\times10^{12}+0.2538$ to match the level of Mg II index in the beginning of 1996. Figure 2 shows a `composite' \citep{Viereck01} Mg II index (dotted line), SEM scaled flux (thin line), and a Running Mean (RM) curve (thick line) with a window of averaging equal to 365 days. The 22/23 minimum determined from the RM (vertical line) corresponds to June 2, 1996.
 \begin{figure}[!ht]
 \plotone{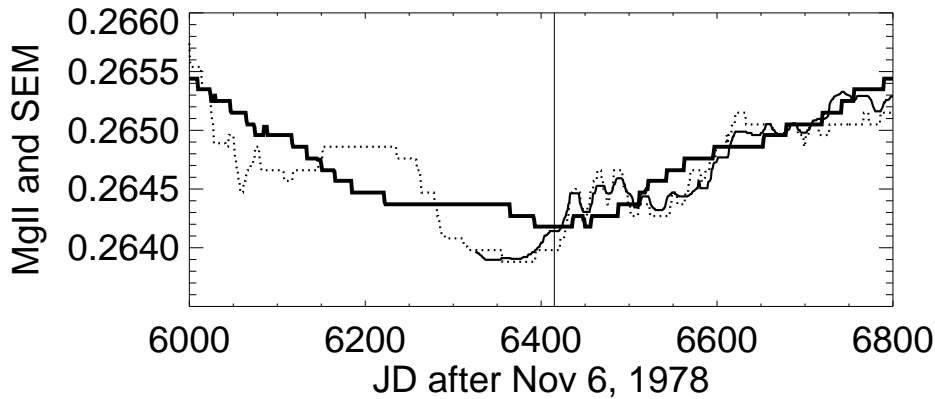}
 \caption{Mg II (dotted line) and SEM scaled flux (thin line) are over-plotted near the 22/23 solar minima. A 365-day RM (thick line) shows the minimum (vertical line) on June 2, 1996.}
 \end{figure}
The scaling factor for SEM flux to match the level of Mg II index in 2008 is $SEM/1.1\times10^{12}+0.255$ (not shown here). This small difference (0.5 \%) between the scaling factors for 22/23 and 23/24 minima confirms the degradation model for SEM is correct and, thus allows us to compare 22/23 and 23/24 minima using SEM vs. SEM absolute EUV flux.

The Mg II index calculated \citep{Snow05} from high spectral resolution (0.1~nm) SORCE/SOLSTICE \citep{McClintock05} observations is available from February 27, 2003. Figure 3 shows a comparison between the SOLSTICE Mg II solar index (top panel, thin line) and SEM scaled flux ($SEM/1.68116\times 10^{11}$, bottom panel, thin line). Demonstrating a significantly different amplitude of variations, both signals show the same local minimum date, July 23, 2008 determined from the RM curves (thick lines) calculated with a 101-day averaging window.
\begin{figure}[!ht]
 \plotone{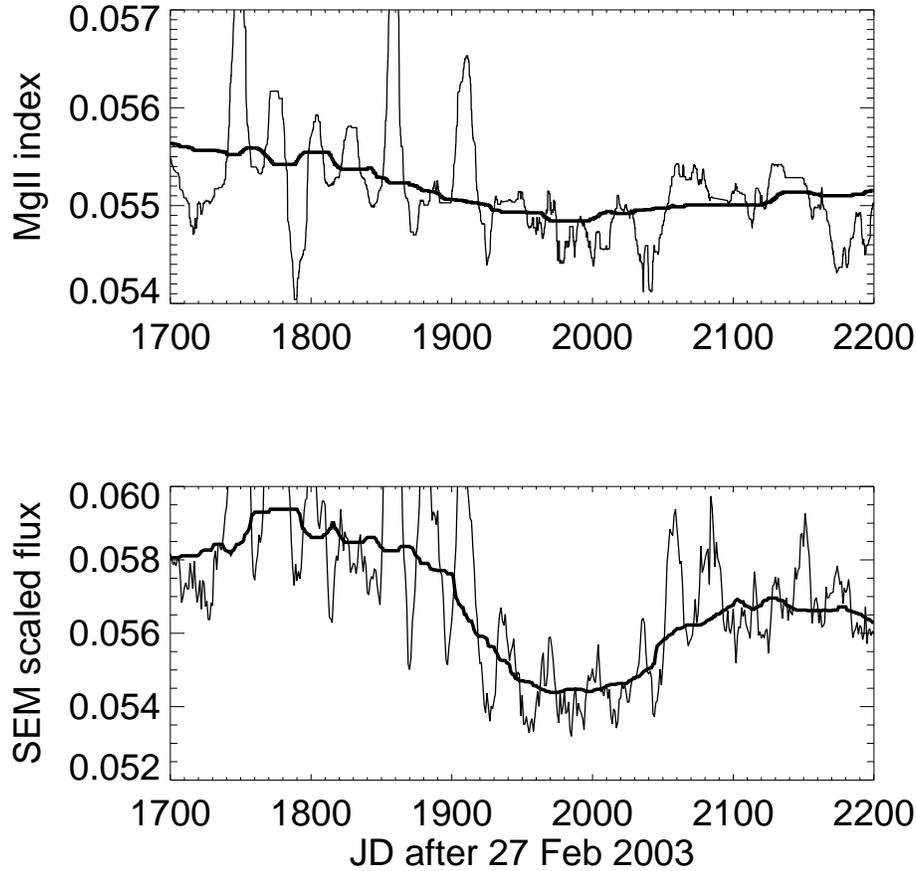}
 \caption{A comparison of SORCE/SOLSTICE Mg II index (top panel, thin line) and SEM scaled flux (bottom panel, thin line) near the 23/24 solar minimum. Thick lines show RM curves with the window of averaging equal to 101 days. The local minima from the RM curves show same date, July 23, 2008.}
 \end{figure}
Figure 4 shows SEM scaled flux ($SEM/1.68116\times 10^{11}$) with about three months (101-day) and one year (365 days) RM curves (dotted and thick lines, respectively). The one year RM is a more realistic estimate of the time of the solar minimum for the solar cycle. It shows that the 23/24 minimum has occurred on November 28, 2008. The SOLSTICE data (not shown in this Figure) give about the same date, November 22, 2008.
\begin{figure}[!ht]
 \plotone{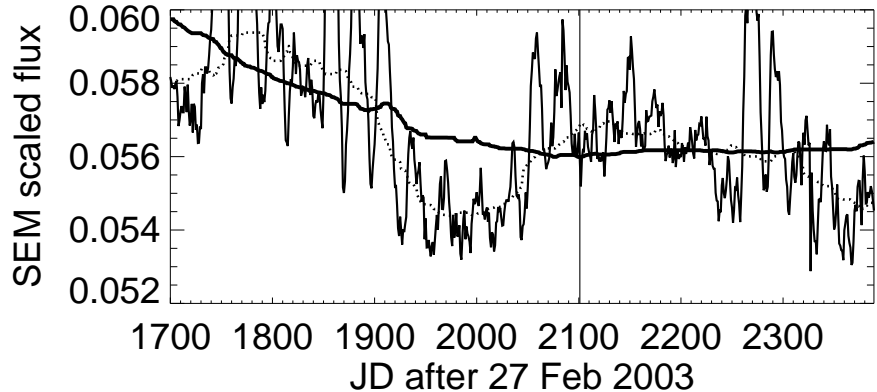}
 \caption{A comparison of solar minima with two RM windows, 101 day (dotted line) and 365 days (solid line) calculated for the scaled SEM flux (thin line). The minimum determined from the RM with one year window (vertical line) is centered on November 28, 2008. }
 \end{figure}

\section{A Comparison of SEM Fluxes for 22/23 and 23/24 Solar Minima}
SEM flux determined from the one-year RM curves for 22/23 and 23/24 minima (Figures 2 and 4) shows a significant decrease of flux level in 2008 compared to 1996, which is about $15\pm 6\%$ for the SEM first order channel centered around the He II (30.4~nm) spectral line. This decreased level is consistent with other observations in different parts of the solar spectrum, see e.g. \citep{Woods09b} or \citep{Lockwood07}. A comparison of the flux RM profiles around the 22/23 and 23/24 solar minima (Figure 2 and 4) shows that the profile (fall and rise) is much less symmetric around the 2008 minimum than it is around the 1996 minimum.

\section{Summary}
The SEM EUV 30.4 nm scaled flux matches the Mg II index for the 22/23 and 23/24 minima. SEM flux shows the minimum for 2008 is lower ($15 \pm 6\%$) than the minimum for 1996, $0.97\times 10^{10} {ph/cm^{2}/s}$ and $1.12\times 10^{10} {ph/cm^{2}/s}$, respectively. This is consistent with recent 04/14/08 SDO/EVE sounding rocket measurements using both EVE/ESP and EVE/MEGS irradiance, and with the result shown by \citet{Lockwood07}.
 The 2008 minimum shows a sharp decrease of irradiance and a steady, slow increase. The 1996 minimum has symmetric wings.
 The 23/24 minima determined from SEM and Mg II data with a RM window of 365-d practically coincide. They occurred on 28 Nov 08 (SEM) and 22 Nov 08 (Mg II).

\acknowledgements This work was partially supported by the NASA grant NNX08AM94G. We thank Marty Snow at LASP for providing the Mg II composite index data.


\begin{thebibliography}{}
\bibitem[Didkovsky et al.(2009)]{D09}
Didkovsky et al. 2009, Sol.Phys., accepted

\bibitem[Hovestadt et al.(1995)]{Hovestadt95}
Hovestadt et al. 1995, Sol.Phys.,162, 441

\bibitem[Judge et al.(1998)]{Judge98}
Judge et al. 1998, Sol.Phys., 177, 161

\bibitem[Judge et al.(2008)]{Judge08}
Judge et al. 2008, AGU Fall Meeting (San Francisco)

\bibitem[Lockwood \& Fr\"{o}hlich(2007)]{Lockwood07}
Lockwood \& Fr\"{o}hlich 2007, Proc. R. Soc. A., 1880

\bibitem[McClintock, Rottman, \& Woods(2007)]{McClintock05}
McClintock, W.E., Rottman, G.J., \& Woods, T.N. 2005, Sol.Phys., 230, 225

\bibitem[Schattenburg \& Anderson(1990)]{Schattenburg90}
Schattenburg, M.L. \& Anderson, E.H. 1990, Phys Scripta 41, 13

\bibitem[Snow et al.(2005)]{Snow05}
Snow et al. 2005, Sol.Phys., 230, 235

\bibitem[Viereck et al.(2001)]{Viereck01}
Viereck et al. 2001, Geophys.Res.Lett., 28,1343

\bibitem[Woods et al.(1998)]{Woods98}
Woods et al. 1998, in Solar Electromagnetic Radiation Study for Solar Cycle 22, ed. J.M. Pap, C. Fr\"{o}hlich and R.K. Ulrich, Sol.Phys., 511

\bibitem[Woods et al.(2009a)]{Woods09a}
Woods et al. 2009a, Sol.Phys., submitted

\bibitem[Woods (2009b)]{Woods09b}
Woods 2009b, ASP Conf. Ser., This Volume, submitted

\end{thebibliography}
\end{document}